\documentclass[lettersize,journal]{IEEEtran}
\usepackage{amsmath,amsfonts}
\usepackage{algorithmic}
\usepackage{algorithm}
\usepackage{multirow}
\usepackage[table,xcdraw]{xcolor}
\usepackage{amsfonts}
\usepackage{array}
\usepackage{comment}
\usepackage{cleveref}
\usepackage{textcomp}
\usepackage{stfloats}
\usepackage{url}
\usepackage{verbatim}
\usepackage{graphicx}
\usepackage{subfigure}
\usepackage{cite}
\begin{document}

\title{TripCEAiR: A Multi-Loss minimization approach for surface EMG based Airwriting Recognition}
\author{Ayush Tripathi, Prathosh A.P., Suriya Prakash Muthukrishnan, and Lalan Kumar

\thanks{This work was supported in part by Prime Minister’s Research Fellowship (PMRF), Ministry of Education (MoE), Government of India.}
\thanks{This work involved human subjects or animals in its research. Approval
of all ethical and experimental procedures and protocols was granted by
the Institute Ethics Committee, All India Institute of Medical Sciences, New Delhi, India with reference number IEC-267/01.04.2022,RP-55/2022.}
\thanks{Ayush Tripathi is with the Department of Electrical Engineering, Indian Institute of Technology Delhi, New Delhi - 110016, India(e-mail: ayush.tripathi@ee.iitd.ac.in).}
\thanks{Prathosh A.P. is with the Department of Electrical Communication Engineering, Indian Institute of Science, Bengaluru - 560012, India(e-mail: prathosh@iisc.ac.in).}
\thanks{Suriya Prakash Muthukrishnan is with the Department of Physiology, All India Institute of Medical Sciences, New Delhi - 110016, India(e-mail: dr.suriyaprakash@aiims.edu).}
\thanks{Lalan Kumar is with the Department of Electrical Engineering,
Bharti School of Telecommunication, and,
Yardi School of Artificial Intelligence, Indian Institute of Technology Delhi, New Delhi - 110016, India(e-mail: lkumar@ee.iitd.ac.in).}
}

% The paper headers
\markboth{Submitted to Elsevier Biomedical Signal Processing and Control}%
{}

% Remember, if you use this you must call \IEEEpubidadjcol in the second
% column for its text to clear the IEEEpubid mark.

\maketitle

\begin{abstract}

Airwriting Recognition refers to the problem of identification of letters written in space with movement of the finger. It can be seen as a special case of dynamic gesture recognition wherein the set of gestures are letters in a particular language. Surface Electromyography (sEMG) is a non-invasive approach used to capture electrical signals generated as a result of contraction and relaxation of the muscles. sEMG has been widely adopted for gesture recognition applications. Unlike static gestures, dynamic gestures are user-friendly and can be used as a method for input with applications in Human Computer Interaction. There has been limited work in recognition of dynamic gestures such as airwriting, using sEMG signals and forms the core of the current work. In this work, a multi-loss minimization framework for sEMG based airwriting recognition is proposed. The proposed framework aims at learning a feature embedding vector that minimizes the triplet loss, while simultaneously learning the parameters of a classifier head to recognize corresponding alphabets. The proposed method is validated on a dataset recorded in the lab comprising of sEMG signals from $50$ participants writing English uppercase alphabets. The effect of different variations of triplet loss, triplet mining strategies and feature embedding dimension is also presented. The best-achieved accuracy was $81.26\%$ and $65.62\%$ in user-dependent and independent scenarios respectively by using semihard positive and hard negative triplet mining. The code for our implementation will be made available at https://github.com/ayushayt/TripCEAiR.

\end{abstract}
\section{Introduction}

\subsection{Background}

The ability to communicate is one of the most important of all life skills that humans possess. The rapid emergence of digital devices has led to a proportional increment in Human Computer Interaction (HCI). However, the medium of user input to HCI systems is limited to traditional methods such as touchscreen, keyboard, and mouse. Therefore, there is a growing demand of alternate HCI input modalities to reduce the need for such additional devices. In this regard, airwriting recognition seems to be a viable solution. Airwriting is referred to as the task of writing in space with the movement of the finger \cite{7322243,7322267}. The unrestricted nature of free space writing provides the user with a seamless method to provide input for HCI applications. Additionally, since the gesture vocabulary is same as that of letters in a language, a user is not required to learn any new gestures for using such a system. Airwriting Recognition has been tackled by using different methods, such as wrist-worn Inertial Measurement Unit \cite{tripathi2021sclair,tripathi2022imair}, smartphone \cite{li2018deep}, wearable glove \cite{amma2013airwriting}, Wii remote \cite{xu2016air}, finger ring \cite{jing2017wearable}, and computer vision based methods \cite{choudhury2021cnn}. In this work, a surface Electromyography (sEMG) based airwriting recognition framework is proposed. sEMG is a physiological signal recorded by placing electrodes on the skin over the target muscle \cite{gohel2020review}. Due to its user-friendly and non-invasive nature, sEMG has been widely used for tasks such as sign language recognition \cite{9747385}, user authentication \cite{9115220}, human machine interaction \cite{simao2019review}, and prosthetic control \cite{samuel2018pattern}. 

\subsection{Related Work}

Gesture recognition using sEMG signals has been used for several applications such as sign language recognition \cite{tateno2020development}, user authentication \cite{9745921}, robot control \cite{boru2022novel}, and rehabilitation \cite{zhou2021toward}. The literature on sEMG based hand gesture recognition can be divided into either static or dynamic gesture recognition. The focus of static gesture recognition is to identify the gestures formed by specified hand shape with no temporal dimension. Several attempts including handcrafted features with machine learning \cite{simao2019review}, and deep learning \cite{li2021gesture} have been proposed for development of static gesture recognition systems. Additionally, several time-domain features \cite{8768831,8558109}, frequency-domain features \cite{7759384} and image representations \cite{8630679,oh2021classification,ozdemir2022hand} of sEMG signals have also been utilized for solving the task of static gesture recognition. In case of dynamic gesture recognition, the task is to identify gestures formed by motion of the hand in space. This adds a temporal dimension to the gestures and thus it becomes essential to simultaneously consider the shape, position and movement of the hand for accurate prediction of the gesture. A CNN-based model trained on time-frequency images was proposed in \cite{9658997} for identification of $5$ dynamic hand gestures. In \cite{yang2021dynamic}, the authors proposed a multi-stream residual network for $6$-class dynamic hand gesture recognition system using sEMG signals. A specific use case of dynamic gesture recognition is the task of handwriting recognition, where the vocabulary of dynamic gestures is the alphabets in a language. A dynamic time warping based approach for handwriting recognition was proposed in \cite{5627246}, and further improvised in \cite{li2013improvements}. A CNN-LSTM based framework for classifying $36$ different gestures, which included $26$ uppercase English alphabets and digits $0$-$9$ was proposed in \cite{beltran2020multi}. Unlike handwriting, the task of airwriting recognition is aimed at identifying characters written in free space, without any visual and haptic feedback during the process of writing. In literature, various airwriting recognition systems have been proposed by using inertial sensors \cite{yanay2020air,tripathi2022imair,tripathi2021sclair} and computer vision based techniques \cite{kim2021writing,MUKHERJEE2019217}. However, there has been limited work in recognition of airwriting by using sEMG signals. To the best of the author's knowledge, an airwriting recognition based on sEMG signals was first proposed in \cite{surfmyoair}. Different time-domain features were utilized to construct sEMG envelopes along with time-frequency images to form input to several deep learning based models for the task of airwriting recognition on a dataset collected in the lab (SurfMyoAiR). Motivated by this, the current study explores a multi-loss minimization approach for airwriting recognition from sEMG signals obtained from a user's forearm muscles.

\subsection{Objectives and Contributions}

In this work, a multi-loss minimization framework for sEMG based airwriting recognition is proposed. The central idea behind the approach is to simultaneously learn feature embeddings from sEMG signals and identify the character corresponding to the given input signal. An encoder block is used to extract embeddings from the input sEMG signal, the parameters of which are learnt by minimizing the triplet loss. The intuition behind triplet loss is to attract the embeddings of an anchor sample close to the embeddings of another sample from the same class (positive sample), while repelling it away from that of a different class (negative sample). These embedding vectors are simultaneously fed to a classifier head, the parameters of which are learnt by minimizing the cross entropy loss for classifying the embeddings into one of the $26$ classes (corresponding to English uppercase alphabets). The entire model is trained in an end-to-end manner by minimizing the sum of triplet and cross entropy losses. The performance of the proposed algorithm is evaluated by performing both user-independent and user-dependent $5$-fold validation on a dataset comprising of sEMG recordings from $50$ participants while writing uppercase English alphabets, collected in the lab. The evaluation is performed by taking three different variants of the triplet loss. Additionally, the variation of recognition accuracies with respect to feature embedding dimension and the choice of triplet mining strategies has been comprehensively explored.

\section{Methodology}

\begin{figure}[t!]
  \centering
  \centerline{\includegraphics[width=\linewidth]{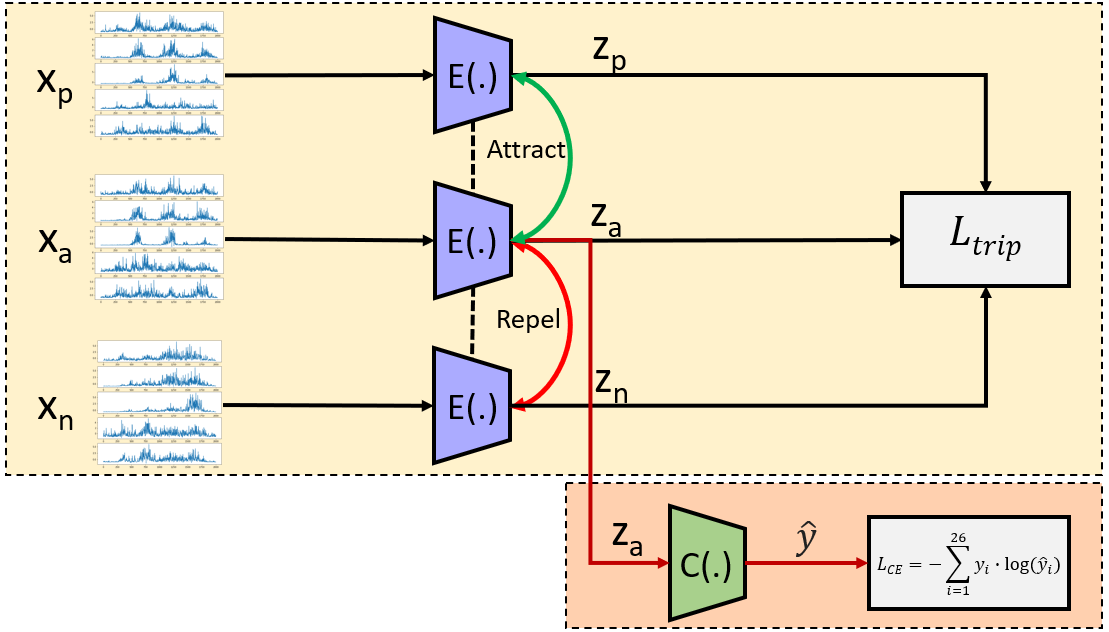}}
  \caption{Block diagram depicting the proposed method. Encoder and classifer blocks are represented by $E(\cdot)$ and $C(\cdot)$ respectively. The model parameters are optimized by minimizing $L_{trip}$ and $L_{CE}$ simultaneously in an end-to-end manner. The dashed lines indicate that the encoders have shared weights. $x_a$, $x_p$, and $x_n$ denote the anchor, positive, and negative samples respectively.} 
\label{fig:blockdiagram}
\end{figure}

The proposed TripCEAiR framework, depicted in Figure \ref{fig:blockdiagram} comprises of two sub-networks: an encoder and a classifier head. The encoder network is used to obtain the feature embeddings by minimizing the triplet loss. The embeddings are subsequently used as input to the classifier network, the parameters of which are learnt by cross-entropy loss minimization. The model is trained in an end-to-end fashion to learn the parameters of both the encoder and the classifier.

\subsection{Proposed Framework}

The input to the model is a multivariate time series comprising of processed sEMG signals recorded while writing English uppercase alphabets. The sEMG signals are obtained from $5$ different locations on the forearm, thereby resulting in a total of five time series per sample. A batch of size $N$ is represented as $\{x_j,y_j\}_{j=1,2,..N}$, where $x_j \in \mathcal{X}$ is the input multivariate time series, and $y_j \in \{A,...Z\}$ is the corresponding alphabet label. The encoder network represented by $E(.): \mathcal{X} \rightarrow \mathbb{R}^{|E|}$, is used to map the input $x_j$ to an embedding vector $z_j$. The parameters of the encoder network denoted by $\theta_E$, are learnt by minimizing the triplet loss ($L_{trip}$) over the embedding vectors after normalizing them to a unit hypersphere (represented by $\Tilde{z_j}$). Minimizing the triplet loss aims at bringing the anchor embedding close to the corresponding positive embedding, while simultaneously pushing it away from the negative embedding. In particular, three different variations of triplet loss are utilized for learning the encoder parameters. 

\subsubsection{N-pair based triplet loss}

A modification of the standard N-pair loss \cite{sohn2016improved} is utilized for learning the feature embeddings. The loss aims at pulling the positive and anchor close to each other while repelling the anchor and negative embeddings by using dot product as a measure of similarity between the embedding vectors. Mathematically, it is given as,

\begin{equation}
   L_{trip-NP} =  \sum_{T(\Tilde{z_a})}^{} - log \frac{exp(\frac{\Tilde{z_a} \cdot \Tilde{z_p} }{\tau})}{exp(\frac{\Tilde{z_a} \cdot \Tilde{z_p} }{\tau}) + exp(\frac{\Tilde{z_a} \cdot \Tilde{z_n} }{\tau})}
\end{equation}
where $T(\Tilde{z_a})$ denotes the set of all possible triplets in a batch, $\tau$ is a scalar parameter (referred to as temperature) and $z_1 \cdot z_2$ represents the inner product between embedding vectors $z_1$ and $z_2$. 

\subsubsection{Margin based triplet loss with $L^2$ norm}

The intuition behind this loss is to minimize the distance between the anchor and positive embeddings while maximizing the distance between the anchor and negative embeddings. Therefore, the objective is to learn embeddings such that, $||\Tilde{z_a} - \Tilde{z_n}||_2^2$ $\geq$ $||\Tilde{z_a} - \Tilde{z_p}||_2^2 + \alpha$. Here, $||\cdot||_2^2$ denotes the square of $L^2$ norm and $\alpha$ is referred to as the margin parameter. To achieve this objective, the loss function is defined as,

\begin{equation}
L_{trip-L2} =  
    \begin{cases}
    0 ,& \text{if } d_{ap} + \alpha \leq d_{an}\\
    d_{ap} - d_{an} + \alpha,              & \text{otherwise}
\end{cases}
\end{equation}

In the equation above, $d_{ap} = ||\Tilde{z_a} - \Tilde{z_p}||_2^2$, and $d_{an} = ||\Tilde{z_a} - \Tilde{z_n}||_2^2$. {By minimizing this loss function, the distance between embeddings of anchor and positive pairs is minimized only when it is greater than the distance between the corresponding anchor and negative pair by a factor $\alpha$. For triplets where this margin is not violated, the loss is set to $0$.} More specifically, the current study uses a smooth version of the aforementioned loss, given by: 

\begin{equation}
   L_{trip-L2} =  \sum_{T(\Tilde{z_a})}^{} log[1 + exp\{d_{ap} - d_{an} + \alpha\}]
\end{equation} 
where, $T(\Tilde{z_a})$ represents the set of all possible triplets within a batch.

\subsubsection{Margin based triplet loss with cosine similarity}

Similar to the margin based loss with $L^2$ norm, the intuition behind this loss is to maximize the similarity between anchor and positive embeddings, while minimizing the similarity between anchor and negative embeddings. The similarity is measured using cosine distance as the metric. In particular, the embeddings are learnt such that $\Tilde{z_a}\cdot\Tilde{z_p}$ $\geq$ $\Tilde{z_a}\cdot\Tilde{z_n} + \alpha$. This objective is achieved by defining the loss function as,

\begin{equation}
L_{trip-CD} =  
    \begin{cases}
    0 ,& \text{if } s_{an} + \alpha \leq s_{ap}\\
    s_{an} - s_{ap} + \alpha,              & \text{otherwise}
\end{cases}
\end{equation}
where, $s_{ap} = \Tilde{z_a}\cdot\Tilde{z_p}$, and $s_{an} = \Tilde{z_a}\cdot\Tilde{z_n}$. {The intuition behind this loss is similar to that of the margin based loss with $L^2$ norm, with the difference lying in the measure of similarity.} With $T(\Tilde{z_a})$ representing the set of triplets in the batch, the smoothed loss function used in the current study is defined as,

\begin{equation}
   L_{trip-CD} =  \sum_{T(\Tilde{z_a})}^{} log[1 + exp\{s_{an} - s_{ap} + \alpha\}]
\end{equation}

The embeddings obtained as output of the encoder block ($z_j$) are simultaneously fed to a classifier network, $C(\cdot)$ which is a mapping from the embedding space to the the set of alphabets. The output of the classifier is a $26$ dimensional vector given by $\hat{y} = C(z) = \sigma(\theta_C^Tz)$. Here, $\theta_C$ are the classifier parameters, which are learnt by minimizing the cross-entropy loss ($L_{CE}$) and $\sigma(\cdot)$ denotes the softmax activation function. The parameters of both encoder and classifier networks are learnt in an end-to-end fashion by minimizing the sum of the two individual losses. Therefore, the final loss to be minimized is the linear combination of the two losses

\begin{equation}
    L = L_{trip} + L_{CE}
\end{equation}

It is to be noted that the triplet loss does not add any additional parameters and hence, during inference, size of the model is same as that of a model trained by only minimizing the cross entropy loss.

\subsection{Triplet Mining Strategies}

Given a batch of $N$ training samples with the input multivariate time series denoted by $x$ and the corresponding label $y_x$, the embedding vector is computed as $z = E(x)$. For an anchor sample $x_a$ (with embedding $z_a$), the corresponding positive and negative are denoted as $x_p$ and $x_n$ (with embeddings $z_p$ and $z_n$) respectively. In this work, several positive and negative mining strategies for forming the triplets are explored \cite{9093432,7298682}. These different strategies are detailed in the following subsections.

\subsubsection{Easy positive mining}

The sample $x_{EP}$ within a batch, the embedding of which is closest to that of the anchor and belongs to the same class is referred to as an easy positive. Mathematically easy positive samples are identified as, 
    
    \begin{equation}
        x_{EP} = \mathop {\operatorname{argmin} }\limits_{\mathop {x:y_x=y_{x_a}} } {||z_a - z||_2}
    \end{equation}
Easy positive mining pulls only the embeddings of two closest positives towards each other. This helps in reducing overclustering and improves generalization of embeddings.
    
\subsubsection{Hard positive mining}

Hard positives ($x_{HP}$) are the samples belonging to the same class as the anchor within the batch that have least similarity with the anchor embeddings. 

\begin{equation}
         x_{HP} = \mathop {\operatorname{argmax} }\limits_{\mathop {x:y_x=y_{x_a}} } {||z_a - z||_2}
\end{equation}
Such a positive mining strategy leads to tight clustering of similar classes in the embedding space. This leads to a decrease in variance and may converge to local minima. 

\subsubsection{Semihard positive mining}

Semihard positive mining aims to select the positive sample from the batch such that, in the embedding space, it is closer from the anchor than the selected negative sample (based on the negative mining strategy). Mathematically, it is represented as,
\begin{equation}
        {x_{SHP}} = \mathop {\operatorname{argmax} }\limits_{\mathop {y_x = y_{x_a}}\limits_{x:||z_a-z||_2  <  ||z_a-z_n||_2} } {||z_a - z||_2}
\end{equation}
Such a mining strategy mitigates the issues of convergence to local minima and possibility of a collapsed model as in case of hard positive mining, while still ensuring tight clustering of the samples belonging to the same class \cite{7298682}. 

\subsubsection{Negative Mining}

Analogous to the positive mining easy, hard, and semihard negatives correspond to samples selected from the batch that belong to a different class from the anchor. Mathematically, it is defined as,

\begin{equation}
    x_{EN} = \mathop {\operatorname{argmax} }\limits_{\mathop {x:y_x \ne y_{x_a}} } {||z_a - z||_2}
\end{equation}
\begin{equation}
         x_{HN} = \mathop {\operatorname{argmin} }\limits_{\mathop {x:y_x \ne y_{x_a}} } {||z_a - z||_2}
\end{equation}
\begin{equation}
        {x_{SHN}} = \mathop {\operatorname{argmin} }\limits_{\mathop {y_x \ne y_{x_a}}\limits_{x:||z_a - z||_2 > ||z_a-z_p||_2} } {||z_a - z||_2}
\end{equation}
Theoretically, easy negatives correspond to the most dissimilar sample within the batch. Hard negative is the sample with the closest embedding to that of the anchor but has a different class label. Similarly, semihard negative chooses a negative from the batch in a way that the embeddings of anchor and the negative are farther than the embeddings of the anchor with the selected positive (based on the positive mining strategy).

\begin{table}[!h]
\caption{Details of the proposed 1DCNN-BiLSTM based encoder.}
\centering
\scalebox{0.78}{
\begin{tabular}{cccc}
\hline
\textbf{Layer}      & \textbf{Kernel Size} & \textbf{\# of filters} & \textbf{Layer Parameters}                   \\\hline\hline
BatchNorm         & -                   & -                    & - \\
Conv1D         & 10                   & 128                    & Stride = 1, Activation = ReLU, Zero padding \\
MaxPool1D        & 3                    &                        & Strides=3, No Padding                       \\
Conv1D         & 10                   & 128                    & Stride = 1, Activation = ReLU, Zero padding \\
MaxPool1D        & 3                    &                        & Strides=3, No Padding                       \\
Conv1D         & 10                   & 256                    & Stride = 1, Activation = ReLU, Zero padding \\
MaxPool1D        & 3                    &                        & Strides=3, No Padding                       \\
Conv1D         & 10                   & 256                    & Stride = 1, Activation = ReLU, Zero padding \\
MaxPool1D        & 3                    &                        & Strides=3, No Padding                       \\
BiLSTM & -                    & -                      & Hidden states = 512, Activation = tanh      \\
Dense               & -                    & -                      & Neurons = $|E|$, Activation = ReLU    \\\hline       
\end{tabular}
}
\label{tab:encoder}
\end{table}

\begin{figure*}[t!]
  \centering
  \centerline{\includegraphics[width=\linewidth]{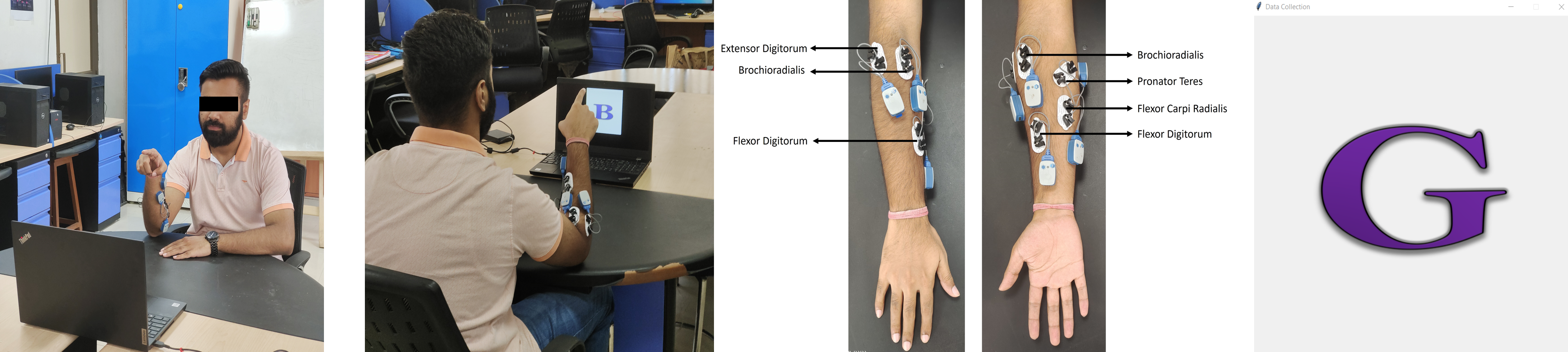}}
  \caption{{Depiction of the sEMG signal recording setup and the electrode placement locations, and the visual stimulus presented to the participant. The figures have been adapted from \cite{surfmyoair}.}}
\label{fig:datacollection}
\end{figure*}

\subsection{Model Architecture}

For the encoder block, an architecture based on a combination of Convolutional Neural Network (CNN) and Bidirectional Long Short Term Memory (BiLSTM) is used. The intuition behind selecting this particular architecture is that the convolutional layers extract the spatial features from the processed sEMG signals. The BiLSTM layer then aims at learning the temporal dimension from these spatial features. The details of the encoder architecture are presented in Table \ref{tab:encoder}. First, Batch Normalization is applied to the processed signals, which are further passed through $4$ convolutional layers each with a kernel size of $10$. For the first couple of convolution layers, the numbers of kernels is set to $128$, while it is increased to $256$ for the last two layers. Maxpooling with a pool size of $3$ is performed after each convolution layer. The features are then fed to a BiLSTM layer with $512$ units. Rectified Linear Unit ($ReLU$) activation function is used for all the convolution layers and hyperbolic tangent ($tanh$) for the BiLSTM layer. The output of the 1DCNN-BiLSTM is subsequently fed to a dense layer comprising of $|E|$ neurons and activated by ReLU activation function to obtain the embedding vector. Further, the classifier head is taken to be a single fully-connected layer having 26 neurons. A dropout of $50\%$ is used in order to avoid overfitting.

\section{Experiments and Results}

\subsection{Dataset Description}

The recording of sEMG signals was done as per guidelines laid down in the Helsinki Declaration, and was ethically approved by the Institute Ethics Committee of the All India Institute of Medical Sciences, New Delhi. The dataset comprises of sEMG signals recorded from five forearm muscles (Flexor Carpi Radialis, Pronator teres, Flexor Digitorum, Brachioradialis, and Extensor Digitorum) while writing English uppercase alphabets ($10$ times). A total of $50$ healthy subjects with mean age of $23.12$ years participated in the experiment with written consent. In order to record the sEMG signals, Noraxon Ultium wireless sEMG sensor \cite{noraxon} and gel-based, self-adhesive Ag/AgCl disposable dual electrodes were used. The sEMG signals were recorded at a sampling rate of $2$ kHz. During the processing stage, the recorded signals were downsampled to $500$Hz and absolute value of the signals was retained. A user interface operated by the experimenter was used to provide visual cue to the participant for the alphabet to be written. Random shuffling of alphabets within a set was done and the participant was provided adequate rest after $2$ repetitions of the alphabet set. {The data collection setup, location of sEMG electrodes on the forearm, and a sample visual stimulus are presented in Figure \ref{fig:datacollection}}.

\begin{table}[t]
\caption{Recognition accuracies for the user-independent airwriting recognition task by using different triplet mining strategies and feature embedding dimension. Entries in blue denote best achieved accuracy for each embedding dimension and the best overall accuracy is depicted in bold.}
\centering
\scalebox{0.68}{
\begin{tabular}{cccccccc}
\hline 
\textbf{Loss}                           & \multicolumn{2}{c}{\textbf{Mining Strategy}} & \multicolumn{5}{c}{\textbf{Embedding Dimension}}                                                                                                                       \\\hline 
                                        & \textbf{Positive}     & \textbf{Negative}    & \textbf{32}                   & \textbf{64}                   & \textbf{128}                  & \textbf{256}                           & \textbf{512}                  \\\hline \hline 
\textbf{$L_{CE}$}                            & \textbf{-}            & \textbf{-}           & 0.6325                        & 0.6345                        & 0.6322                        & 0.6278                                 & 0.6340                        \\\hline 
                                        & Easy                  & Easy                 & 0.6277                        & 0.6417                        & 0.6367                        & 0.6409                                 & 0.6361                        \\
                                        & Easy                  & Hard                 & 0.6294                        & 0.6376                        & 0.6412                        & 0.6490                                 & 0.6457                        \\
                                        & Easy                  & Semihard             & 0.6394                        & 0.6456                        & 0.6492                        & 0.6433                                 & 0.6462                        \\
                                        & Easy                  & All                  & 0.6339                        & 0.6378                        & 0.6445                        & 0.6447                                 & 0.6411                        \\
                                        & Hard                  & Easy                 & 0.6298                        & 0.6292                        & 0.6358                        & 0.6294                                 & 0.6302                        \\
                                        & Hard                  & Hard                 & 0.6083                        & 0.6172                        & 0.6195                        & 0.6221                                 & 0.6188                        \\
                                        & Hard                  & Semihard             & 0.6372                        & 0.6465                        & 0.6356                        & 0.6432                                 & 0.6488                        \\
                                        & Hard                  & All                  & 0.6228                        & 0.6164                        & 0.6266                        & 0.6301                                 & 0.6259                        \\
                                        & Semihard              & Easy                 & 0.6306                        & 0.6340                        & 0.6363                        & 0.6243                                 & 0.6295                        \\
                                        & Semihard              & Hard                 & {\color[HTML]{0000FF} 0.6462} & {\color[HTML]{0000FF} 0.6483} & {\color[HTML]{0000FF} 0.6546} & {\color[HTML]{0000FF} \textbf{0.6562}} & {\color[HTML]{0000FF} 0.6553} \\
                                        & All                   & Easy                 & 0.6272                        & 0.6348                        & 0.6287                        & 0.6354                                 & 0.6347                        \\
                                        & All                   & Hard                 & 0.6095                        & 0.6365                        & 0.6383                        & 0.6331                                 & 0.6375                        \\
\multirow{-13}{*}{\textbf{$L_{trip-NP} + L_{CE}$}} & All                   & All                  & 0.6423                        & 0.6342                        & 0.6394                        & 0.6422                                 & 0.6466                        \\\hline 
                                        & Easy                  & Easy                 & 0.6048                        & 0.6226                        & 0.6286                        & 0.6263                                 & 0.6175                        \\
                                        & Easy                  & Hard                 & 0.6393                        & 0.6355                        & 0.6323                        & 0.6307                                 & 0.6314                        \\
                                        & Easy                  & Semihard             & 0.6358                        & 0.6388                        & 0.6354                        & 0.6355                                 & 0.6295                        \\
                                        & Easy                  & All                  & 0.6233                        & 0.6280                        & 0.6313                        & 0.6387                                 & 0.6405                        \\
                                        & Hard                  & Easy                 & 0.6066                        & 0.6155                        & 0.6251                        & 0.6223                                 & 0.6202                        \\
                                        & Hard                  & Hard                 & 0.6248                        & 0.6176                        & 0.6219                        & 0.6272                                 & 0.6246                        \\
                                        & Hard                  & Semihard             & 0.6282                        & 0.6336                        & 0.6317                        & 0.6369                                 & 0.6332                        \\
                                        & Hard                  & All                  & 0.6267                        & 0.6308                        & 0.6356                        & 0.6376                                 & 0.6327                        \\
                                        & Semihard              & Easy                 & 0.6022                        & 0.6158                        & 0.6227                        & 0.6245                                 & 0.6151                        \\
                                        & Semihard              & Hard                 & 0.6305                        & 0.6351                        & 0.6328                        & 0.6244                                 & 0.6372                        \\
                                        & All                   & Easy                 & 0.6092                        & 0.6302                        & 0.6255                        & 0.6287                                 & 0.6262                        \\
                                        & All                   & Hard                 & 0.6307                        & 0.6225                        & 0.6310                        & 0.6412                                 & 0.6279                        \\
\multirow{-13}{*}{\textbf{$L_{trip-CD} + L_{CE}$}} & All                   & All                  & 0.6308                        & 0.6403                        & 0.6420                        & 0.6417                                 & 0.6355                        \\\hline 
                                        & Easy                  & Easy                 & 0.6236                        & 0.6222                        & 0.6328                        & 0.6298                                 & 0.6251                        \\
                                        & Easy                  & Hard                 & 0.6241                        & 0.6328                        & 0.6333                        & 0.6415                                 & 0.6393                        \\
                                        & Easy                  & Semihard             & 0.6412                        & 0.6392                        & 0.6388                        & 0.6286                                 & 0.6261                        \\
                                        & Easy                  & All                  & 0.6410                        & 0.6399                        & 0.6391                        & 0.6431                                 & 0.6397                        \\
                                        & Hard                  & Easy                 & 0.6158                        & 0.6272                        & 0.6217                        & 0.6261                                 & 0.6210                        \\
                                        & Hard                  & Hard                 & 0.6126                        & 0.6285                        & 0.6198                        & 0.6314                                 & 0.6273                        \\
                                        & Hard                  & Semihard             & 0.6327                        & 0.6431                        & 0.6364                        & 0.6362                                 & 0.6412                        \\
                                        & Hard                  & All                  & 0.6211                        & 0.6358                        & 0.6347                        & 0.6359                                 & 0.6377                        \\
                                        & Semihard              & Easy                 & 0.6125                        & 0.6193                        & 0.6252                        & 0.6145                                 & 0.6169                        \\
                                        & Semihard              & Hard                 & 0.6424                        & 0.6394                        & 0.6495                        & 0.6402                                 & 0.6367                        \\
                                        & All                   & Easy                 & 0.6262                        & 0.6179                        & 0.6307                        & 0.6218                                 & 0.6275                        \\
                                        & All                   & Hard                 & 0.6112                        & 0.6395                        & 0.6356                        & 0.6273                                 & 0.6332                        \\
\multirow{-13}{*}{\textbf{$L_{trip-L2} + L_{CE}$}} & All                   & All                  & 0.6292                        & 0.6420                        & 0.6437                        & 0.6412                                 & 0.6419   \\ \hline                    
\end{tabular}
}
\label{tab:Results_userindep}
\end{table}

\subsection{Experimental Details}

The recorded sEMG signals correspond to different letters from different users, thereby leading to a variation in the length of signals. To mitigate this, the absolute sEMG signals are interpolated to a length of $L$ samples using cubic interpolation if the signal was shorter and the extra samples were discarded, otherwise. The length $L$ was taken to be $2000$ samples which corresponds to $4$ second duration of writing.  Subsequently, z-normalization was performed to each of the processed signals individually. 

In order to check for robustness of the proposed method, experiments were performed in both user-independent and user-dependent settings. In the user-independent approach, $5$-fold validation was performed while ensuring no subject overlap in the training and test sets. More specifically, in each fold, data corresponding to $40$ subjects was used for training the model and evaluation was performed on the held-out subjects. Similarly, in the user-dependent setting, $5$-fold validation was performed by splitting the entire data based on the repetition number during the airwriting data collection process. In each fold, the model was trained on $8$ repetitions of all the subjects, and tested on the left out $2$ repetitions. The training data was further split in an $80:20$ ratio to form the training and validation sets. A mini-batch training process with a batch size of $260$ was employed and the parameters of the model were updated using the Adam optimizer. Early stopping with a patience of $10$ epochs while monitoring the validation accuracy was employed to avoid overfitting of the model. 

\begin{table}[t]
\caption{Recognition accuracies for the user-dependent airwriting recognition task by using different triplet mining strategies and feature embedding dimension. Entries in blue denote best achieved accuracy for each embedding dimension and the best overall accuracy is depicted in bold.}
\centering
\scalebox{0.68}{
\begin{tabular}{cccccccc}
\hline 
\textbf{Loss}                           & \multicolumn{2}{c}{\textbf{Mining Strategy}} & \multicolumn{5}{c}{\textbf{Embedding Dimension}}                                                                                                                       \\\hline 
                                        & \textbf{Positive}     & \textbf{Negative}    & \textbf{32}                   & \textbf{64}                   & \textbf{128}                  & \textbf{256}                           & \textbf{512}                  \\\hline \hline 
\textbf{$L_{CE}$}                           & \textbf{-}            & \textbf{-}           & 0.7549                        & 0.7712                        & 0.7748                        & 0.7725                                 & 0.7701                        \\\hline 
                                        & Easy                  & Easy                 & 0.7583                        & 0.7664                        & 0.7823                        & 0.7764                                 & 0.7815                        \\
                                        & Easy                  & Hard                 & 0.7725                        & 0.7792                        & 0.7840                        & 0.7898                                 & 0.7924                        \\
                                        & Easy                  & Semihard             & 0.7867                        & 0.7999                        & 0.7955                        & 0.7898                                 & 0.7904                        \\
                                        & Easy                  & All                  & 0.7781                        & 0.7761                        & 0.7729                        & 0.7900                                 & 0.7856                        \\
                                        & Hard                  & Easy                 & 0.7484                        & 0.7677                        & 0.7665                        & 0.7772                                 & 0.7718                        \\
                                        & Hard                  & Hard                 & 0.7533                        & 0.7445                        & 0.7684                        & 0.7677                                 & 0.7645                        \\
                                        & Hard                  & Semihard             & 0.7836                        & 0.7843                        & 0.7906                        & 0.7852                                 & 0.7992                        \\
                                        & Hard                  & All                  & 0.7608                        & 0.7829                        & 0.7879                        & 0.7778                                 & 0.7744                        \\
                                        & Semihard              & Easy                 & 0.7555                        & 0.7611                        & 0.7677                        & 0.7705                                 & 0.7745                        \\
                                        & Semihard              & Hard                 & {\color[HTML]{0000FF} 0.7861} & {\color[HTML]{0000FF} 0.8035} & {\color[HTML]{0000FF} 0.8053} & {\color[HTML]{0000FF} \textbf{0.8126}} & {\color[HTML]{0000FF} 0.8089} \\
                                        & All                   & Easy                 & 0.7545                        & 0.7575                        & 0.7641                        & 0.7811                                 & 0.7797                        \\
                                        & All                   & Hard                 & 0.7709                        & 0.7701                        & 0.7819                        & 0.7540                                 & 0.7533                        \\
\multirow{-13}{*}{\textbf{$L_{trip-NP} + L_{CE}$}} & All                   & All                  & 0.7826                        & 0.7879                        & 0.7938                        & 0.7847                                 & 0.7872                        \\\hline 
                                        & Easy                  & Easy                 & 0.7379                        & 0.7505                        & 0.7548                        & 0.7606                                 & 0.7620                        \\
                                        & Easy                  & Hard                 & 0.7518                        & 0.7688                        & 0.7742                        & 0.7746                                 & 0.7657                        \\
                                        & Easy                  & Semihard             & 0.7606                        & 0.7715                        & 0.7675                        & 0.7782                                 & 0.7547                        \\
                                        & Easy                  & All                  & 0.7557                        & 0.7667                        & 0.7795                        & 0.7882                                 & 0.7815                        \\
                                        & Hard                  & Easy                 & 0.7334                        & 0.7571                        & 0.7628                        & 0.7509                                 & 0.7335                        \\
                                        & Hard                  & Hard                 & 0.7558                        & 0.7500                        & 0.7749                        & 0.7645                                 & 0.7717                        \\
                                        & Hard                  & Semihard             & 0.7565                        & 0.7751                        & 0.7655                        & 0.7766                                 & 0.7718                        \\
                                        & Hard                  & All                  & 0.7606                        & 0.7698                        & 0.7724                        & 0.7794                                 & 0.7668                        \\
                                        & Semihard              & Easy                 & 0.7482                        & 0.7413                        & 0.7545                        & 0.7513                                 & 0.7325                        \\
                                        & Semihard              & Hard                 & 0.7843                        & 0.7678                        & 0.7821                        & 0.7788                                 & 0.7712                        \\
                                        & All                   & Easy                 & 0.7428                        & 0.7552                        & 0.7603                        & 0.7680                                 & 0.7660                        \\
                                        & All                   & Hard                 & 0.7610                        & 0.7484                        & 0.7489                        & 0.7675                                 & 0.7685                        \\
\multirow{-13}{*}{\textbf{$L_{trip-CD} + L_{CE}$}} & All                   & All                  & 0.7650                        & 0.7893                        & 0.7822                        & 0.7772                                 & 0.7922                        \\\hline 
                                        & Easy                  & Easy                 & 0.7466                        & 0.7656                        & 0.7673                        & 0.7622                                 & 0.7645                        \\
                                        & Easy                  & Hard                 & 0.7718                        & 0.7689                        & 0.7724                        & 0.7782                                 & 0.7723                        \\
                                        & Easy                  & Semihard             & 0.7697                        & 0.7743                        & 0.7565                        & 0.7874                                 & 0.7658                        \\
                                        & Easy                  & All                  & 0.7550                        & 0.7725                        & 0.7838                        & 0.7887                                 & 0.7908                        \\
                                        & Hard                  & Easy                 & 0.7512                        & 0.7580                        & 0.7688                        & 0.7665                                 & 0.7568                        \\
                                        & Hard                  & Hard                 & 0.7422                        & 0.7672                        & 0.7659                        & 0.7488                                 &  0.7408                             \\
                                        & Hard                  & Semihard             & 0.7692                        & 0.7778                        & 0.7769                        & 0.7808                                 & 0.7952                        \\
                                        & Hard                  & All                  & 0.7500                        & 0.7722                        & 0.7810                        & 0.7815                                 & 0.7702                        \\
                                        & Semihard              & Easy                 & 0.7356                        & 0.7498                        & 0.7572                        & 0.7587                                 & 0.7672                        \\
                                        & Semihard              & Hard                 & 0.7891                        & 0.7778                        & 0.7929                        & 0.7871                                 & 0.7906                        \\
                                        & All                   & Easy                 & 0.7489                        & 0.7572                        & 0.7608                        & 0.7725                                 & 0.7622                        \\
                                        & All                   & Hard                 & 0.7575                        & 0.7604                        & 0.7652                        & 0.7682                                 & 0.7659                        \\
\multirow{-13}{*}{\textbf{$L_{trip-L2} + L_{CE}$}} & All                   & All                  & 0.7682                        & 0.7896                        & 0.7925                        & 0.7892                                 & 0.7894         \\ \hline               
\end{tabular}
}
\label{tab:Results_userdep}
\end{table}

\begin{figure}[t!]
  \centering
  \centerline{\includegraphics[width=0.9\linewidth]{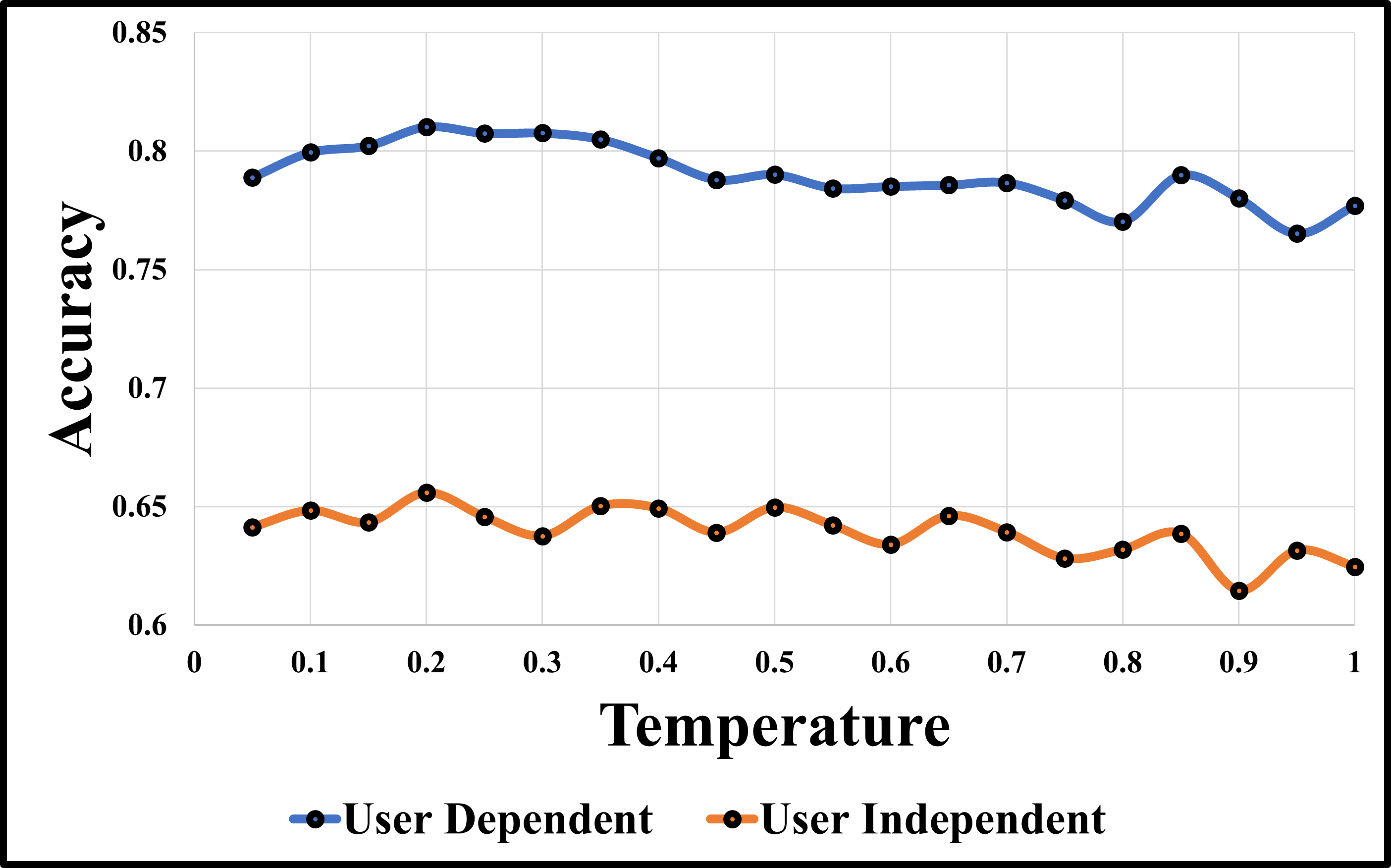}}
  \caption{Effect of temperature parameter on recognition accuracy for sEMG based airwriting recognition in user dependent and independent scenarios.} 
\label{fig:temperature_plot}
\end{figure}

\begin{figure}[t!]
  \centering
  \centerline{\includegraphics[width=0.7\linewidth]{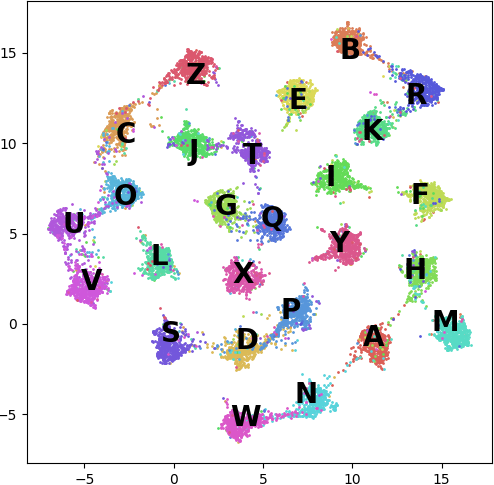}}
  \caption{Scatter plot depicting clusters for the different alphabets in the embedding space.} 
\label{fig:umap_plot}
\end{figure}

\begin{table}[h]
\caption{{Alphabet-wise precision, recall and F1-score corresponding to Semihard positive and Hard negative mining with $L_{trip-NP} + L_{CE}$ loss in user-independent, and user-dependent evaluation settings.}}
\centering
\scalebox{0.85}{
{\begin{tabular}{ccccccc}
\hline 
                  & \multicolumn{3}{c}{\textbf{User Independent}}            & \multicolumn{3}{c}{\textbf{User Dependent}}              \\\hline
\textbf{Alphabet} & \textbf{Precision} & \textbf{Recall} & \textbf{F1-Score} & \textbf{Precision} & \textbf{Recall} & \textbf{F1-Score} \\\hline\hline
\textbf{A}        & 0.73               & 0.74            & 0.74              & 0.87               & 0.84            & 0.86              \\
\textbf{B}        & 0.74               & 0.69            & 0.72              & 0.88               & 0.84            & 0.86              \\
\textbf{C}        & 0.69               & 0.73            & 0.71              & 0.77               & 0.78            & 0.77              \\
\textbf{D}        & 0.54               & 0.51            & 0.52              & 0.68               & 0.67            & 0.67              \\
\textbf{E}        & 0.68               & 0.74            & 0.71              & 0.84               & 0.83            & 0.83              \\
\textbf{F}        & 0.65               & 0.62            & 0.63              & 0.79               & 0.80            & 0.80              \\
\textbf{G}        & 0.70               & 0.66            & 0.68              & 0.84               & 0.82            & 0.83              \\
\textbf{H}        & 0.69               & 0.67            & 0.68              & 0.84               & 0.82            & 0.83              \\
\textbf{I}        & 0.59               & 0.52            & 0.55              & 0.81               & 0.75            & 0.78              \\
\textbf{J}        & 0.64               & 0.59            & 0.61              & 0.82               & 0.82            & 0.82              \\
\textbf{K}        & 0.66               & 0.65            & 0.65              & 0.81               & 0.77            & 0.79              \\
\textbf{L}        & 0.64               & 0.69            & 0.66              & 0.81               & 0.80            & 0.81              \\
\textbf{M}        & 0.74               & 0.76            & 0.75              & 0.87               & 0.90            & 0.88              \\
\textbf{N}        & 0.59               & 0.61            & 0.60              & 0.77               & 0.77            & 0.77              \\
\textbf{O}        & 0.67               & 0.66            & 0.67              & 0.79               & 0.78            & 0.79              \\
\textbf{P}        & 0.57               & 0.56            & 0.56              & 0.71               & 0.70            & 0.71              \\
\textbf{Q}        & 0.65               & 0.63            & 0.64              & 0.82               & 0.83            & 0.83              \\
\textbf{R}        & 0.64               & 0.64            & 0.64              & 0.76               & 0.82            & 0.79              \\
\textbf{S}        & 0.73               & 0.81            & 0.77              & 0.85               & 0.87            & 0.86              \\
\textbf{T}        & 0.56               & 0.55            & 0.56              & 0.78               & 0.78            & 0.78              \\
\textbf{U}        & 0.64               & 0.64            & 0.64              & 0.71               & 0.77            & 0.74              \\
\textbf{V}        & 0.64               & 0.69            & 0.67              & 0.79               & 0.80            & 0.79              \\
\textbf{W}        & 0.67               & 0.72            & 0.69              & 0.81               & 0.82            & 0.82              \\
\textbf{X}        & 0.65               & 0.68            & 0.66              & 0.82               & 0.84            & 0.83              \\
\textbf{Y}        & 0.64               & 0.62            & 0.63              & 0.85               & 0.79            & 0.82              \\
\textbf{Z}        & 0.73               & 0.71            & 0.72              & 0.82               & 0.85            & 0.84   \\\hline          
\end{tabular}
}}
\label{tab:precrecall}
\end{table}

\begin{figure*}[!ht]
\centering  
\subfigure[]{\includegraphics[width=\linewidth]{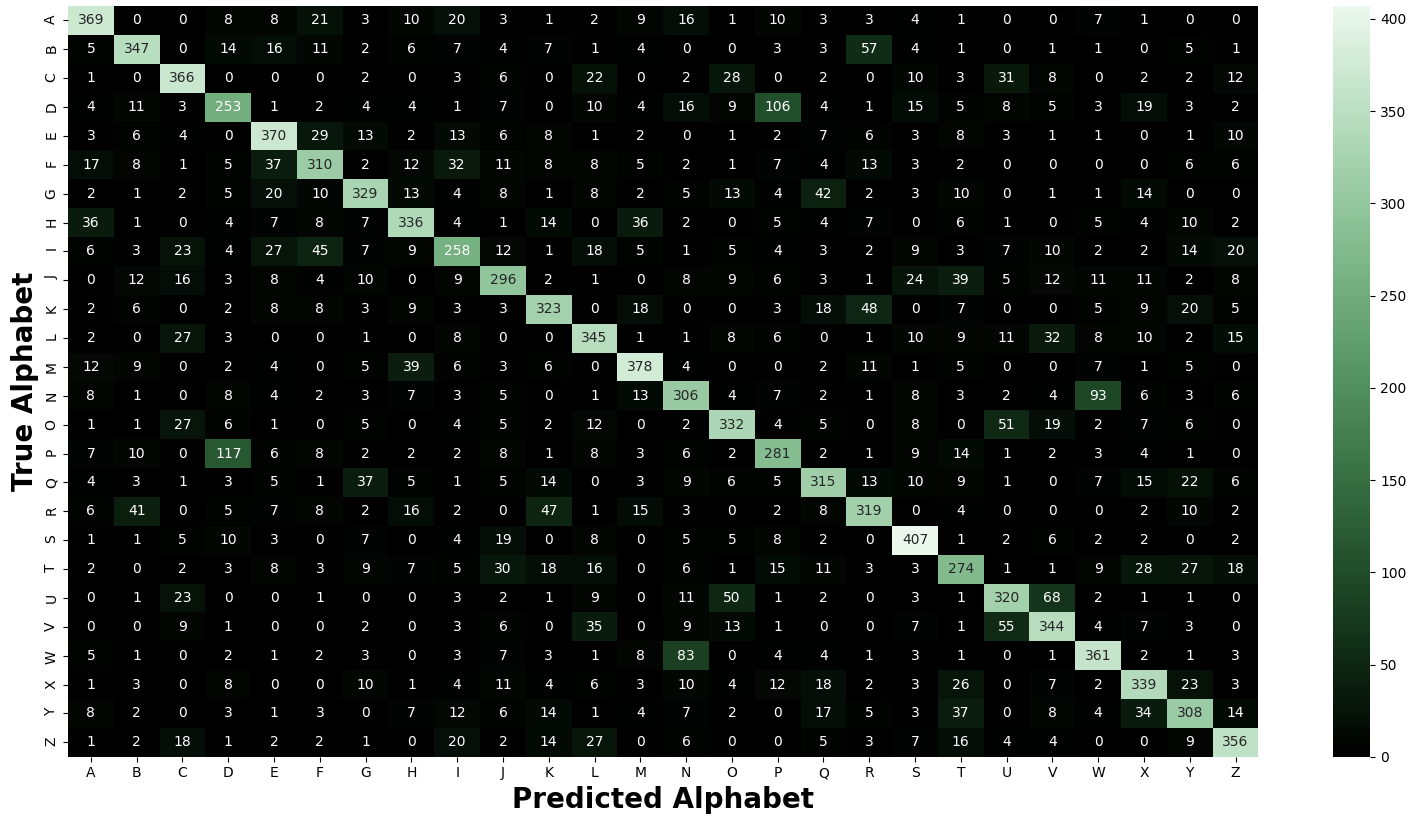}}
\subfigure[]{\includegraphics[width=\linewidth]{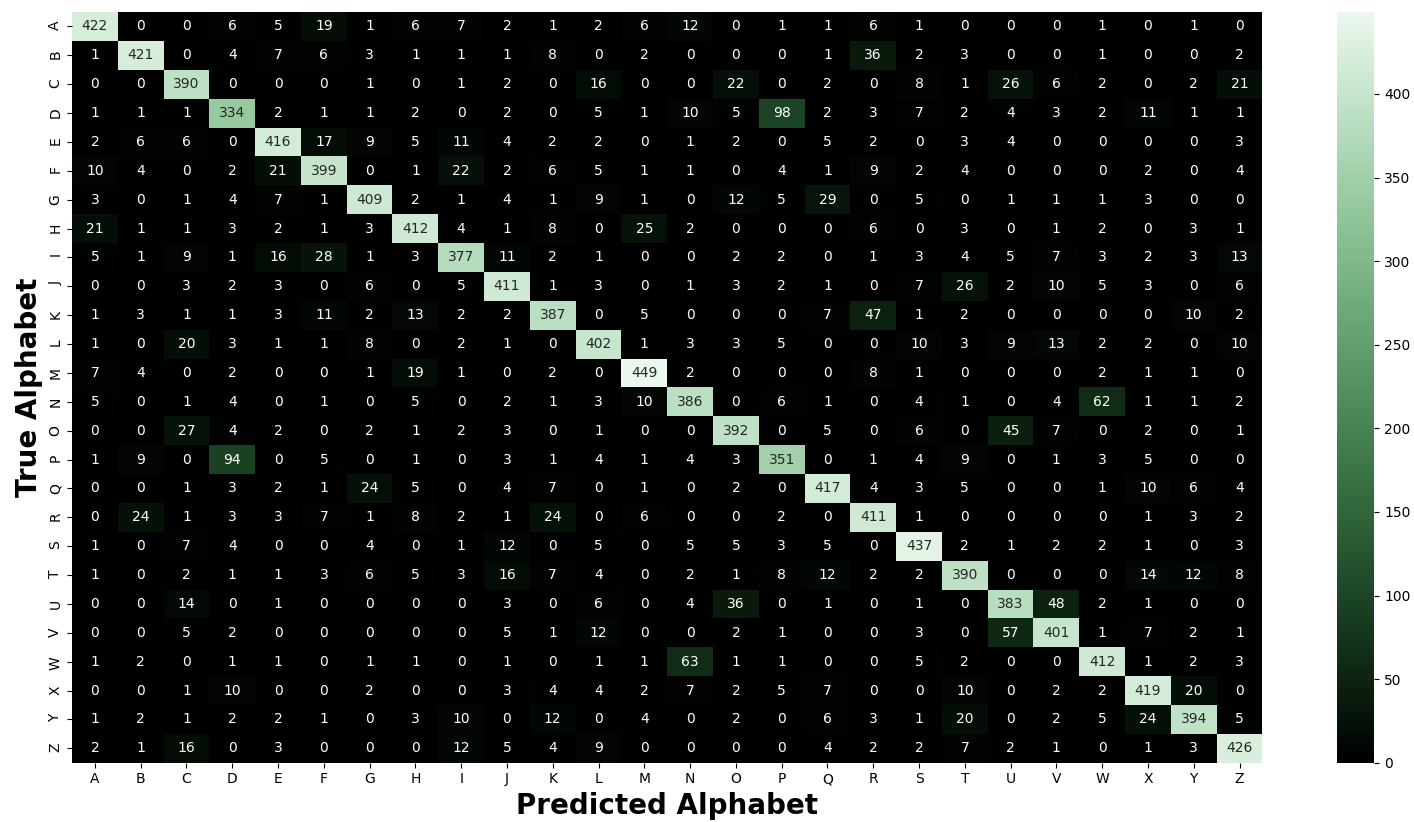}}
\caption{Confusion matrices corresponding to Semihard positive and Hard negative mining with $L_{trip-NP} + L_{CE}$ loss in (a) user-independent, and (b) user-dependent evaluation settings.}
\label{fig:ConfMat}
\end{figure*}

\begin{table*}[t]
\caption{Comparison of different model architectures for sEMG based airwriting recognition task.}
\centering
\scalebox{0.9}{
\begin{tabular}{cccccc}
\hline 
                                  & \textbf{Architecture}   & \textbf{$L_{CE}$}  & \textbf{$L_{trip-NP} + L_{CE}$} & \textbf{$L_{trip-L2} + L_{CE}$} & \textbf{$L_{trip-CD} + L_{CE}$} \\\hline\hline
\multirow{5}{*}{User Dependent}   & \textbf{1DCNN}          & 0.6479       & 0.6522                & 0.6538                & 0.6506                \\
                                  & \textbf{Stacked LSTM}   & 0.4651       & 0.4419                & 0.4383                & 0.4400                \\
                                  & \textbf{Stacked BiLSTM} & 0.4810       & 0.4545                & 0.4394                & 0.4479                \\
                                  & \textbf{1DCNN-LSTM}     & 0.7822       & 0.8089                & 0.7823                & 0.7746                \\
                                  & \textbf{1DCNN-BiLSTM}   & 0.7725       & \textbf{{\color[HTML]{0000FF} 0.8126}}               & 0.7871                & 0.7788                \\\hline\hline
\multirow{5}{*}{User Independent} & \textbf{1DCNN}          & 0.5415       & 0.5432                & 0.5482                & 0.5552                \\
                                  & \textbf{Stacked LSTM}   & 0.3335       & 0.3221                & 0.3049                & 0.3103                \\
                                  & \textbf{Stacked BiLSTM} & 0.4047       & 0.3692                & 0.3475                & 0.3419                \\
                                  & \textbf{1DCNN-LSTM}     & 0.6385       & 0.6559                & 0.6420                & 0.6375                \\
                                  & \textbf{1DCNN-BiLSTM}   & 0.6278       & \textbf{{\color[HTML]{0000FF} 0.6562}}                & 0.6402                & 0.6244 \\\hline              
\end{tabular}
}
\label{tab:comparison}
\end{table*}

\subsection{Results and Discussion}

The effect of variation in mean recognition accuracies with different values of the temperature parameter ($\tau$) for the N-pair based triplet loss are presented in Figure \ref{fig:temperature_plot}. For observing the role of $\tau$, the embedding dimension is set to $256$, along with semihard positive and hard triplet mining. It is observed that the best accuracy is achieved with $\tau = 0.2$, and this value is retained for all subsequent experiments. Similarly, the optimal values of the margin parameter ($\alpha$) in case of margin-based triplet loss is identified. The value is found to be $\alpha = 0.2$ and $\alpha=0.1$ for margin loss with $L^2$ norm and cosine similarity, respectively. Tables \ref{tab:Results_userindep} and \ref{tab:Results_userdep} list the performance of the proposed architecture with the accuracies averaged across the $5$ folds. The variation of accuracy with different feature embedding dimension (i.e. $|E|$), triplet mining strategies and the three variations of the triplet loss used in the study is also presented. It is observed from the tables that using the N-pair based triplet loss yields an improved accuracy compared to the margin based losses. It is seen that the best accuracy is achieved by using a feature embedding dimension of $256$ along with semihard positive and hard negative mining. The superior accuracies obtained by using this triplet mining technique may be attributed to the fact that such a strategy pushes apart the most dissimilar samples while pulling those samples that are closer to the anchor than the hardest negative. {The feature embedding vectors, which are the output of the encoder block (denoted by $z$) are visualized by reducing to $2$ dimensions by using Uniform Manifold Approximation and Projection (UMAP). The plot depicting this visualization is presented in Figure \ref{fig:umap_plot}.} The plot reveals tight clusters corresponding to each of the $26$ uppercase English alphabets, with similar-looking alphabets (such as D \& P) close to each other in the embedding space. The same is also observed from the confusion matrices presented in Figures \ref{fig:ConfMat}. It can be seen that similar-looking alphabets which are close to each other in the embedding space lead to the highest misclassification. {Additionally, the precision, recall and F1-scores for individual alphabets is presented in Table \ref{tab:precrecall}.} Overall, on combining the N-pair based triplet loss, an absolute improvement of $2.84\%$ and $4.01\%$ is observed in case of user independent and dependent scenarios respectively. Furthermore, the performance of the airwriting recognition system by using different model architectures including 1DCNN, Stacked Long Short Term Memory (LSTM), Bidirectional LSTM and 1DCNN-LSTM/BiLSTM is also presented in Table \ref{tab:comparison}. It is observed that the proposed 1DCNN-BiLSTM outperforms the other model architectures. This is attributed to the fact that in this architecture, the CNN layers act as a feature extractor and max pooling layer takes the average of the feature within a sliding window. The BiLSTM layer then learns the temporal characteristics from the extracted features. Hence, this combination of CNN-BiLSTM is able to better capture the nuances of the sEMG signal corresponding to different alphabets leading to superior performance.

\section{Conclusion}

In this paper, a multi-loss framework by combining triplet and cross entropy losses (TripCEAiR) for sEMG based airwriting recognition is proposed. Additionally, the effect of different triplet mining strategies and feature embedding dimension is comprehensively explored. It is seen that the proposed method outperforms the approach wherein model parameters are learnt by minimizing only the cross-entropy loss. An accuracy of $81.26\%$ and $65.62\%$ was obtained in the user-dependent and user-independent scenarios respectively by using semihard positive and hard negative mining and the N-pair based triplet loss. This implies an absolute improvement of $4.01\%$ and $2.84\%$ in the user-dependent and independent settings over the case when only cross-entropy loss is minimized. The improvement in recognition accuracies indicate the usability of the proposed framework in developing an sEMG based airwriting recognition system for HCI applications. {Future work may be focused on exploring the robustness of the proposed sEMG based airwriting recognition system in adverse scenarios such as the presence of excessive noise and imperfect sensor placement.}

\bibliographystyle{IEEEtran}
\bibliography{refs}

\clearpage

\section{Appendix}

Details of the encoder architectures used for comparison are presented in \cref{tab:1dcnn,tab:LSTM,tab:BiLSTM,tab:cnnlstm}.

\begin{table}[ht!]
\caption{Details of the 1DCNN based encoder.}
\label{tab:1dcnn}
\centering
\scalebox{0.75}{
\begin{tabular}{cccc}
\hline
\textbf{Layer}  & \textbf{Kernel Size} & \textbf{\# of filters} & \textbf{Layer Parameters}                   \\\hline\hline
BatchNorm         & -                   & -                    & - \\
Conv1D          & 10                   & 128                    & Stride = 1, Activation = ReLU, Zero padding \\
MaxPool1D       & 3                    &                        & Strides=3, No Padding                       \\
Conv1D          & 10                   & 128                    & Stride = 1, Activation = ReLU, Zero padding \\
MaxPool1D       & 3                    &                        & Strides=3, No Padding                       \\
Conv1D          & 10                   & 256                    & Stride = 1, Activation = ReLU, Zero padding \\
MaxPool1D       & 3                    &                        & Strides=3, No Padding                       \\
Conv1D          & 10                   & 256                    & Stride = 1, Activation = ReLU, Zero padding \\
MaxPool1D       & 3                    &                        & Strides=3, No Padding                       \\
GlobalAvgPool1D & -                    & -                      & -                                           \\
Dense               & -                    & -                      & Neurons = $|E|$, Activation = ReLU    \\\hline
\end{tabular}
}
\end{table}

\begin{table}[ht!]
\caption{Details of the Stacked LSTM-based encoder.}
\label{tab:LSTM}
\centering
\scalebox{0.78}{
\begin{tabular}{cccc}
\hline
\textbf{Layer}  & \textbf{Kernel Size} & \textbf{\# of filters} & \textbf{Layer Parameters}                   \\\hline\hline
BatchNorm         & -                   & -                    & - \\
LSTM           & -                    & -                      & Hidden states = 512, Activation = tanh \\
LSTM           & -                    & -                      & Hidden states = 512, Activation = tanh \\
Dense               & -                    & -                      & Neurons = $|E|$, Activation = ReLU    \\\hline
\end{tabular}
}
\end{table}

\begin{table}[ht!]
\caption{Details of the Stacked Bidirectional LSTM-based encoder.}
\label{tab:BiLSTM}
\centering
\scalebox{0.78}{
\begin{tabular}{cccc}
\hline
\textbf{Layer}  & \textbf{Kernel Size} & \textbf{\# of filters} & \textbf{Layer Parameters}                   \\\hline\hline
BatchNorm         & -                   & -                    & - \\
BiLSTM           & -                    & -                      & Hidden states = 512, Activation = tanh \\
BiLSTM           & -                    & -                      & Hidden states = 512, Activation = tanh \\
Dense               & -                    & -                      & Neurons = $|E|$, Activation = ReLU    \\\hline
\end{tabular}
}
\end{table}

\begin{table}[!h]
\caption{Details of the 1DCNN-LSTM based encoder.}
\centering
\scalebox{0.78}{
\begin{tabular}{cccc}
\hline
\textbf{Layer}      & \textbf{Kernel Size} & \textbf{\# of filters} & \textbf{Layer Parameters}                   \\\hline\hline
BatchNorm         & -                   & -                    & - \\
Conv1D         & 10                   & 128                    & Stride = 1, Activation = ReLU, Zero padding \\
MaxPool1D        & 3                    &                        & Strides=3, No Padding                       \\
Conv1D         & 10                   & 128                    & Stride = 1, Activation = ReLU, Zero padding \\
MaxPool1D        & 3                    &                        & Strides=3, No Padding                       \\
Conv1D         & 10                   & 256                    & Stride = 1, Activation = ReLU, Zero padding \\
MaxPool1D        & 3                    &                        & Strides=3, No Padding                       \\
Conv1D         & 10                   & 256                    & Stride = 1, Activation = ReLU, Zero padding \\
MaxPool1D        & 3                    &                        & Strides=3, No Padding                       \\
LSTM & -                    & -                      & Hidden states = 512, Activation = tanh      \\
Dense               & -                    & -                      & Neurons = $|E|$, Activation = ReLU    \\\hline       
\end{tabular}
}
\label{tab:cnnlstm}
\end{table}

\end{document}